\documentclass[10pt,a4paper,dvips,twocolumn,showpacs]{revtex4}
\usepackage[english]{babel}
\usepackage{epsfig}
\usepackage{graphics, curves} 
\usepackage{amsmath, amsfonts, amssymb}
\usepackage{graphicx}
\usepackage{dcolumn}
\usepackage{dcolumn}
\begin{document}
\title{Anomalous Scaling and Solitary Waves in Systems with 
Non-Linear Diffusion}
\author{Alex Hansen}
\email[Alex.Hansen@ntnu.no]{}
\affiliation{Department of Physics, Norwegian University of Science and
Technology, N--7491 Trondheim, Norway}

\author{Bo-Sture Skagerstam}
\email[Bo-Sture.Skagerstam@ntnu.no]{}
\affiliation{Department of Physics, Norwegian University of Science and
Technology, N--7491 Trondheim, Norway}

\author{Glenn T{\o}r{\aa}}
\email[Glenn.Tora@ntnu.no]{}
\affiliation{Department of Physics, Norwegian University of Science and
Technology, N--7491 Trondheim, Norway}

\begin{abstract}
We study a non-linear convective-diffusive equation, local in space and time, 
which has its background in the dynamics of the thickness of a wetting film.
The presence of a non-linear diffusion predicts the existence of fronts as well 
as shock fronts. Despite the absence of memory effects, solutions in the case 
of pure non-linear diffusion exhibit an anomalous sub-diffusive scaling.
Due to a  balance between non-linear diffusion and convection we, in particular, 
show that solitary waves appear. For large times they merge into a single 
solitary wave exhibiting a topological stability. Even though our results 
concern a specific equation, numerical simulations supports the view that 
anomalous diffusion and the solitary waves disclosed will be  general 
features in such non-linear convective-diffusive dynamics.
\end{abstract}
\maketitle

Anomalous diffusion is a topic of great current interest 
(see e.g.\ Refs.\ \cite{bg90,metzler_2000,engelsberg}).
It is commonly believed that the phenomenon
has its roots in correlated structures and/or disorder in the medium in which the diffusion
occurs which, in turn, induce memory effects.  However, it is the aim of the
present work to demonstrate that a non-linearity in the diffusion 
process, local in space and time, is sufficient to induce anomalous scaling.  
When, in addition,  convection enters the non-linear diffusion problem we consider, we
find, in particular, that the same non-linearities that cause the anomalous diffusion,
also lead to the occurrence of solitary waves exhibiting a topological 
stability.  The topological stability manifests itself through the solitary 
waves having their speed given by their average amplitude.   
     
Even though the non-linear diffusion-convection equation 
that we study has its origins in the 
dynamics of wetting films, 
related equations appear in other context like 
non-linear heat conduction \cite{zeldovich67,mayer83}.  
Wetting films have, in their own right,   
received much attention recently due to their 
relevance for flows in porous media 
(see e.g. Refs.\cite{w21})
and soil physics and hydrology \cite{hall07}.  
Shock-wave features are in general expected in systems with non-linear 
convection and diffusion \cite{whitham_74}.  In elastic media such features 
has recently been observed in the laboratory \cite{fink_2003}.  Due to the 
generic  non-linearities in the equations of motion an exact 
mathematical analysis
is often not possible and  numerical simulations are then expected 
to be  mandatory.  The lack of an exact mathematical analysis may, 
however,  prevent one from a deeper  insight into the physical 
consequences of the non-linear dynamics.  

\begin{figure}[t]
\begin{picture}(0,0)(200,390)   

\includegraphics{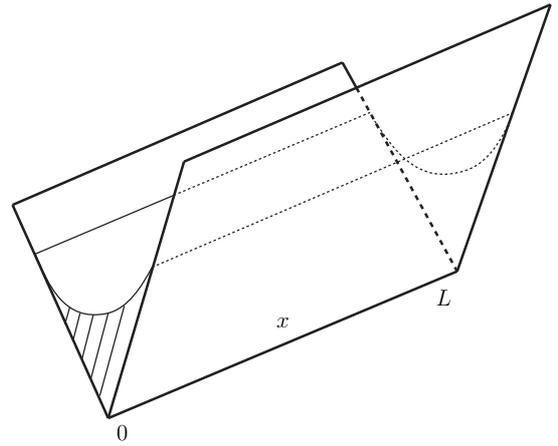}
\end{picture}
\vspace{5.5cm}
\caption{\label{fig_flow} Flow of a wetting film in a wedge. The shaded area denotes a cross-section area $a(x,t)$ of the film which, in general, various along the wedge.}
\end{figure}

 In order to be specific, we consider a capillary pore flow 
with a wetting film flowing in a closed wedge with a length $L$ as illustrated in Fig.\ref{fig_flow}
The geometry chosen appears not be essential but enables us to derive 
specific and analytical predictions. We 
denote the position along the wedge with  $x$ such that $0 \le x \le L$, 
and where, e.g., the endpoints  could be connected to other pores. 
For reasons of simplicity we will consider an infinite half-line and hence 
make $L$ arbitrarily large.  The cross sectional area of the 
film at $x$ is $a\equiv a(x,t)$, where $t \ge 0$ is the time variable.  The 
total and fixed cross sectional area of the wedge is
$A_T$ so that the remaining cross section is $\delta A(x,t)=A_T-a$.  
Apart from the
wetting film, the closed wedge is filled with a non-wetting fluid.  There
is a pressure drop $P\equiv P(x,t)$ along the wedge in the non-wetting fluid 
and a pressure drop $p\equiv p(x,t)$ in the wetting film.  The non-wetting 
fluid flows with 
a volume flow rate $Q\equiv Q(x,t)$ and the wetting fluid with a volume 
flow rate $q \equiv q(x,t)$ so 
that the total $Q_T=Q+q$ is constant.  Both fluids are assumed to be 
incompressible.  
The non-wetting fluid has a viscosity $M$ and the wetting fluid a 
viscosity $\mu$.    

Conservation of mass dictates that the area $a$ is related to the flow
rate through the equation
\begin{equation}
\label{wetcons}
\frac{\partial q}{\partial x}+\frac{\partial a}{\partial t}=0\;.
\end{equation}
We consider a flow in the 
films governed by the Darcy equation \cite{d92}, i.e.
\begin{equation}
\label{filmdarcy}
q = -\frac{k(a)}{\mu}\ \frac{\partial p}{\partial x}\;.
\end{equation}
Here $k(a)=a^2/8\pi$ is the effective and $a$-dependent 
film permeability of the film, assuming a Hagen-Poiseuille 
flow \cite{d92}.
The Young relation now gives the connection between pressure difference across
the interface between wetting and non-wetting fluids and the curvature
of the interface $r$ at capillary equilibrium \cite{d92}, i.e. $P-p=\gamma /r$, 
where $\gamma$ is the surface tension.  If the wedge has
an opening angle  $\phi$ and the wetting angle is zero, then 
a geometrical consideration leads to 
$a=r^2 \left( \cot\left(\phi /2\right)-\phi/ 2 \right ) $.
Combining this relation with the Young relation leads to 
\begin{equation}
\label{pvsa}
p=P-\frac{\overline{\gamma}}{\sqrt{a}}\;\;,
\end{equation} 
where we have defined the effective surface tension 
$\overline{\gamma}=\gamma \sqrt{a}/r$.
If we now eliminate $p$ and $q$ between Eqs.\ (\ref{wetcons}) --
(\ref{pvsa}), we find our basic equation of motion
\begin{equation}
\label{diffeq_1}
\frac{\partial}{\partial x} \left[ a^2\ \frac{\partial}{\partial x}
\left(P-\frac{\overline{\gamma}}{\sqrt{a}}\right)\right] = 8\pi\mu\
\frac{\partial a}{\partial t}\;.
\end{equation}

We now make the assumption that the pressure in the non-wetting
fluid, $P$, is unaffected by the area $a$ of the film.  This is
true if $\delta A>>a$. The Darcy equation (\ref{filmdarcy}) then reads
$MQ/K=-\partial P /\partial x$, 
where  $K=A_T^2/8\pi$ is the effective permeability of  
an assumed Hagen-Poiseuille bulk flow.
Due to conservation of mass of the non-wetting fluid we have
$\partial Q/\partial x = 0$, 
which, after a straightforward integration, leads to
\begin{equation}
\label{intnonwetdarcy}
\frac{\partial P(x,t)}{\partial x}=
- \frac{8M\pi {\bar Q}}{A_T^2}\;,
\end{equation}
where the pressure drop at $x=0$ is neglected, i.e., 
$\partial P(0,t)/\partial x = 0$, and ${\bar Q} \simeq Q>0$ 
is a characteristic volume flow rate. All time 
dependencies enter through the non-wetting pressure $P$ at $x=0$. 
If there is no flow
of non-wetting fluid in the pore, 
$Q=0$,  the pressure $P$ is constant along it.

The two fundamental
equations in the present paper for the evolution of the film in 
the wedge are Eqs.\ (\ref{diffeq_1}) and (\ref{intnonwetdarcy}).  We non-dimensionalize
and combine them by setting $a={\bar a} \alpha$, $x={\bar x} \xi$ and 
$t={\bar \tau}\tau$,  where ${\bar a},{\bar x}, {\bar \tau}$ are 
some typical area, length and time-scales.  
The combined equation is the {\it diffusion-convection equation\/}
\begin{equation}
\label{diffeq_2}
\frac{\partial}{\partial\xi}\left( \ w \sqrt{\alpha}
\frac{\partial\alpha}{\partial\xi}\right) 
- v \alpha \frac{\partial\alpha}{\partial\xi}
= \frac{\partial\alpha}{\partial\tau}\;.
\end{equation}
where $w={\bar a}^{1/2} {\bar \tau} \overline{\gamma}/(16\pi\mu {\bar x}^2)$ 
and $v=2M{\bar a}{\bar \tau}{\bar Q}M/(A_T^2{\bar x} \mu)$ 
can be regarded to be the relevant  
physical parameters in the system.  The diffusivity in Eq.\ (\ref{diffeq_2})  
is $D=w\sqrt{\alpha}$ and there is also a convective flow 
velocity $v\alpha $ both of which therefore are now non-trivial
functions of $a$. 

 The effective diffusivity $D=w\sqrt{\alpha}$ is counter
intuitive when interpreting the diffusion process literary: The diffusion
coefficient increases with increasing concentration contrary to 
conventional diffusion where higher concentration means a more crowded system and,
hence, slower diffusion.
The non-linear convective term  $v \alpha \partial\alpha/\partial\xi$ in
Eq.~(\ref{diffeq_2}) is well known from surface wave propagation in shallow
water (see e.g.\ Ref. \cite{whitham_74}).  It corresponds to a 
wave moving towards increasing $\xi$ with a
speed proportional to $\alpha$.  In the context of shallow-water theory, it
is the term that produces breaking waves: An initially smooth wave will
eventually produce a singularity at the leading side.  

When there is no 
non-wetting fluid flow, only the diffusive term is present on the left
hand side of Eq.\ (\ref{diffeq_2}). 
 Eq.~(\ref{diffeq_2}) without convection was considered Ref.\cite{dong95}  as a model for experimental results of imbibition processes. 
The experimental presence of an imbibition front was theoretically 
inferred using approximative solutions of Eq.\ (\ref{diffeq_2}) 
\cite{mayer83}. 
Here we point out the existence of such fronts as well 
as shock fronts as exact consequences of Eq.\ (\ref{diffeq_2}). 
It is also a remarkable consequence of our analysis that anomalous 
scaling laws  of the form $\xi ^2 \simeq t^{\gamma} $ with $\gamma \neq1$  
can be obtained without invoking fractional derivatives 
\cite{metzler_2000} in diffusion-like 
equations (see e.g. Refs.\ \cite{engelsberg}) or
by referring to stochastic Mori-Lee equations \cite{mori_lee}  and memory 
effects (see e.g.\ Refs.\ \cite{morgado_2002,schutz_2004}).
If the effective diffusivity $D$ is an increasing (or decreasing) function of $\alpha $
we find sub-diffusion with $\gamma < 1$ (or super-diffusion with $\gamma > 1$). 
A related observation of the appearance of anomalous diffusion  in a class of non-linear Fokker-Planck equations can be found in Ref.\cite{tsallis_96}. A renormalization group analysis of the so called  Barenblatt's non-linear diffusion equation without convection, which describes the filtration of a certain compressible fluid through an elastic porous medium, also leads to anomalous diffusion \cite{liu_90}. The diffusive non-linearity has then, however,  a completely different origin and form as compared to the non-linearities considered in the present paper.

The  non-linear diffusion equation obtained from  Eq.~(\ref{diffeq_2}) with $v=0$ can now be transformed 
from a partial differential equation to an ordinary one 
in terms of a scaling function $h$ such that 
$\alpha(\xi,\tau )=h\left(\xi/f(\tau)\right)/f(\tau)$.  
This choice of  scaling of the solution is not unique but is, 
by construction, such that a load $E$ defined as 
$E=\int_0^{L/\overline{x}} d\xi \alpha(\xi,\tau )$ is conserved in 
time provided it is finite.  We now obtain
\begin{equation}
\label{alpha_0}
h_0=\sqrt{h}\ \frac{dh}{dy} + cyh \;.
\end{equation} 
as well as 
\begin{equation}
\label{func}
f(\tau) =(1+\frac{5cw\tau}{2})^{2/5} \;,
\end{equation} 
in terms of two constants, $h_0$ and $c$,  of integration.
Here we have defined the natural Boltzmann scaling variable $y =\xi/f(\tau)$ 
\cite{crank75}.  We notice that the scaling form of the solution
and Eq.\ (\ref{func}) leads to anomalous scaling since $h\simeq constant$
corresponds to $\xi ^2 \simeq \tau^{4/5}$, i.e. sub-diffusion, and that  
solutions of the scaling form $\alpha(\xi,\tau)= h\left(\xi/g(\tau)\right)$ lead to
the conventional scaling $g(\tau) \simeq \tau^{1/2}$.
The constant of integration  $h _0$ in Eq.\ (\ref{alpha_0})
can have either sign.  The constant $c>0$ can be 
determined from a condition on 
$\partial \alpha(\xi,\tau )/ \partial \tau$ at $\tau=0$ or by 
the load $E$. As long as $h_0 \neq 0$ we can now write 
the solution $\alpha(\xi,\tau )$ in the form
\begin{equation}
\label{scaling}
\alpha(\xi,\tau ) = \frac{|h_0|}{(|h_0|c^2)^{1/5}f(t)} \tilde{h}(\tilde{y}) \;,
\end{equation} 
where we have defined 
$\tilde{y} =\xi c/(|h_0|c^2)^{1/5}f(\tau)$ and 
where $\tilde{h} \equiv \tilde{h}(\tilde{y})$ now obeys the equation
\begin{equation}
\label{scaled_h}
k=\sqrt{\tilde{h}}\ \frac{d\tilde{h}}{d\tilde{y}} + \tilde{y}\tilde{h} \;,
\end{equation} 
with $k = h_0/|h_0| = \pm 1$. 
The scale transformations 
$\xi\rightarrow \lambda \xi ;\tau \rightarrow \lambda ^2 \tau; 
c \rightarrow c/\lambda ^2$ and $|h_0| \rightarrow |h_0|/\lambda$ 
leaves $\tilde{y}$ and $\tilde{h}$ invariant and hence also 
all solutions $\alpha(\xi,\tau )$ in the form as given by Eq.\ (\ref{scaling}). 
 Let us now assume that the film is concentrated initially
around the origin, $\xi =0$.  Hence, $d\alpha /dy <0$, and we must have
that $k=-1$.  Eq.~(\ref{scaled_h}) 
then shows that the solution  $\tilde{h}$ flows 
towards $\tilde{h}(y_{\max})=0$ at some finite value $y=y_{\max}$ 
with an infinite derivative, i.e. we have a shock front. 
Even though we do not, at present, have an analytical result for $y_{\max}$ 
given $\tilde{h}(0)$,  one can easily calculate $y_{\max}$ numerically.  
We then find that $y_{\max} \simeq \log \tilde{h}(0)$, at least 
for large values of $\tilde{h}(0)$.  Since $y_{\max}$ is finite 
the load $E$ will also be finite, which then  
leads to a possible determination of $c$ in terms of $E$. If, on the other hand, 
$k=1$ we find, using Eq.\ (\ref{scaled_h}), 
that $\tilde{h}$ will have a maximum and then flows to zero 
asymptotically according to $\tilde{h} \simeq 1/\tilde{y}$. In this case 
the load $E$ is not well defined and the constant of integration 
$c$ can then be determined using initial data as mentioned above.
\begin{figure}[]
\leftline{\includegraphics [width=0.95\columnwidth]{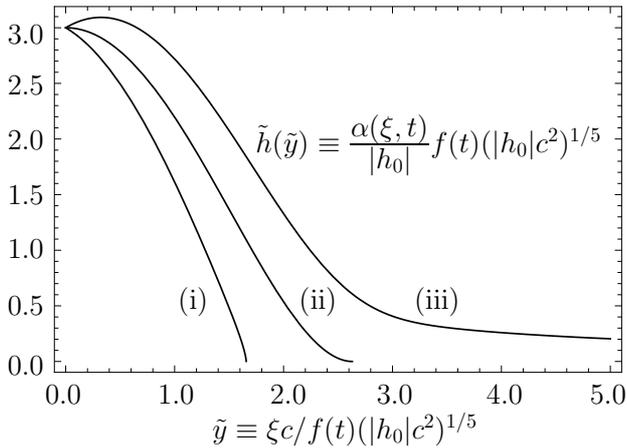}}
\caption{\label{fig3-1} Summarizing, with $\tilde{h}(0)= 3$, 
the three branches of
the integral curves Eq.\ (\ref{scaled_h}) 
with $k=-1$ (curve (i) with a shock front),  
and $k=1$ (curve (iii) with a front). 
In the case $k=0$ (curve (ii)) we make 
the replacement $|h_0| \rightarrow \tilde{h}(0)^{5/4}$. }
\end{figure}
Finally, we have to consider the special case with $k=0$ in Eq.\ 
(\ref{scaled_h}) which then can be solved explicitly. 
In the scaling variable $\tilde{y}$ one then makes the replacement 
$|h_0| \rightarrow \tilde{h}(0)^{5/4}$ and we can then write
\begin{equation}
\label{zel_sol}
\alpha (\xi ,\tau) = 
\frac{\alpha _0}{f(t)} 
\left( 1- \frac{\xi ^2 c}{4f(\tau)\alpha _0^{1/2}}  \right) ^2 \; , 
\end{equation}
where $\alpha_0 \equiv \alpha(0,0)$. In this case the 
constant of integration $c$ can, e.g., be determined by the 
conserved load $E$. In passing, we notice that the exact solution Eq.\ (\ref{zel_sol}) 
corresponds to the known  Zeldovich-Raizer solution in the 
context of non-linear heat conduction \cite{zeldovich67,mayer83}.
We show the various scaling solutions 
$\tilde{h}(\tilde{y})$ of Eq.\ (\ref{scaled_h})  graphically in
Fig.\ \ref{fig3-1} for a specific initial value.  
As we have shown above  the solutions with $k=-1$ and 
$k=0$ have then a finite range.
All solutions exhibits anomalous scaling due to Eq.\ (\ref{func}).

 We will now turn to the existence of solitary wave propagation in the pore. 
In the following we solve the film equation Eq.\ (\ref{diffeq_2}) 
under the
assumption that the system is infinitely long.  
Hence, we make a shift of the origin $x=0$ and let $L \rightarrow \infty$ so
that $\xi$ can take any value on $(-\infty,+\infty)$.
Intuitively, when, in a non-linear system, the convective term is non-zero, there may be propagating solutions  where the convective term and the diffusive
terms balance each other as is the case for the Burgers equation \cite{whitham_74}. The dynamical equation  we consider, i.e. Eq.(\ref{diffeq_2}),  actually has the same convective term as the Burgers equation but describes the diffusion in a non-linear manner.  In our case, the primary physical reason for the balance alluded to above therefore is
the effective diffusion constant which is proportional to $\sqrt{\alpha}$
and hence increases the diffusion rate with increasing $\alpha$.

By defining the natural variable $z=\xi-U\tau\;$ with $U > 0$ , 
then, after an integration, Eq.~(\ref{diffeq_2}) is transformed into
\begin{equation}
\label{integraK}
K=w\sqrt{\alpha}\ \left(\frac{d\alpha}{dz}\right)-
\left(\frac{v}{2}\right)\ \alpha^2
+U\ \alpha\;, 
\end{equation}
where $K$ is an integration constant. We now rewrite this equation in the form
\begin{equation}
\label{split}
\frac{\sqrt{\alpha}}{(\alpha-\alpha_+)(\alpha-\alpha_-)}\frac{d\alpha}{dz} 
= \frac{v}{2w} \; ,
\end{equation}
where we notice that $\alpha_+ \alpha_- = 2K/v$
and that the velocity $U$ of the solitary wave is given by
\begin{equation}
\label{velocity}
 \alpha_++\alpha_- = \frac{2U}{v}\; ,
\end{equation}
i.e.,
\begin{equation}
\label{sing}
\alpha_\pm
=\left(\frac{U}{v}\right)\ 
\left(1\pm\sqrt{1-\frac{2Kv}{U^2}}\right)\, .
\end{equation}
We now search for solitary like solutions $\alpha_s \equiv \alpha$ to Eq.\ (\ref{split}) 
whose variations are localized
in space, i.e., $d\alpha_s/d\xi\to 0$ as $|\xi|$ becomes large or, 
equivalently in terms of the variable $z$, as  
$\tau \to\pm\infty$. In these limits,
the integration constant $K$ defined in Eq.\ (\ref{integraK}) 
becomes $K=U\alpha_\pm - v\alpha_\pm^2/2$,
where we  chose $\alpha_{+}> \alpha_{-}\ge 0$ for the 
two asymptotic areas of the film and 
hence $U>0$. Since $\alpha_\pm$ are real
 we conclude from Eq.\ (\ref{sing}) 
that $0 \le K = v\alpha_+\alpha_- /2< U^2/2v$. 
Eq.\ (\ref{integraK}) can now be integrated in a straightforward manner, i.e., 
\begin{eqnarray}
\label{soliton_3}
\sqrt{\alpha_-}\log\left(\frac{\sqrt{\alpha_s}+\sqrt{\alpha_-}}{\sqrt{\alpha_s}-
\sqrt{\alpha_-}}\right)- \sqrt{\alpha_+}
\log\left(\frac{\sqrt{\alpha_s}+\sqrt{\alpha_+}}{\sqrt{\alpha_+}
-\sqrt{\alpha_s}}\right)
 =\nonumber\\
\frac{v}{2w}(\alpha_+ -\alpha_-)\left(z-2z_0\right)\; , ~~~~~~~~~~~~~~
\end{eqnarray}
where $z_0$ constitutes the initial condition.  
We note that $\alpha_s\to\alpha_+$ as 
$\tau \to \infty$ and $\alpha_s \to \alpha_-$ as 
$\tau \to -\infty$ as long as the 
initial conditions are chosen such that 
$\alpha_+>\alpha_s(z_0)>\alpha_-$. The left-hand 
side of Eq.\ (\ref{soliton_3}) now 
describes a $\alpha_s$-profile that is translated in time. 
With a given initial data $\alpha_s(z_0)$ the asymptotic 
values $\alpha_{\pm}$ are finite. It is a 
straightforward calculation to show that the 
solution $\alpha_s$ of the form Eq.\ (\ref{soliton_3}) 
is stable  for small deformations at large values of $|\xi|$. 
This follows from the observation that any solution 
$\alpha = \alpha_s + \delta$  of   Eq.\ (\ref{diffeq_2}) 
leads to a diffusion equation for 
$\delta$ keeping terms to linear order in 
$\delta$ for large values of $|\xi|$.

\begin{figure}[]
\vspace{6mm}
\leftline{\includegraphics [width=0.95\columnwidth]{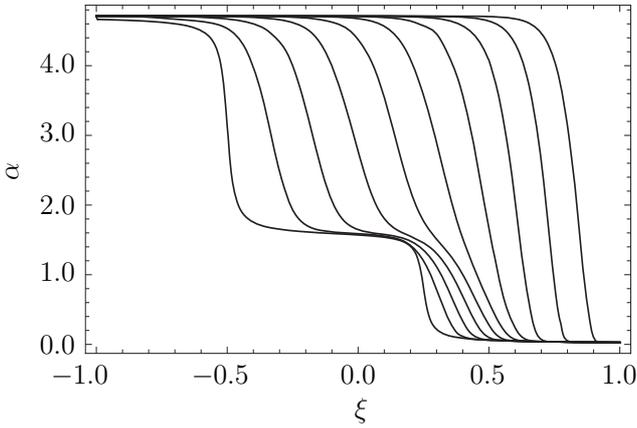}}
\caption{\label{fig7-2} FEM solution of Eq.\ (\ref{diffeq_2}) 
with $w=1.0$, $v=2.0$ with
an initial two-step configuration  
$\alpha(\xi,\tau=0)=[\pi/2-\arctan(50\xi+25)]+[\pi/2-\arctan(50\xi-25/2)]/2$.
The subsequent evolution of this profile is shown at intervals of 
$\Delta\tau=0.025$. The velocity of the profiles at later times,  
corresponding to the solitary wave Eq.\ (\ref{soliton_3}), 
fulfills Eq.\ (\ref{velocity}) in the main text. 
}
\end{figure}
\begin{figure}[]
\vspace{1mm}
\leftline{\includegraphics [width=1.0\columnwidth]{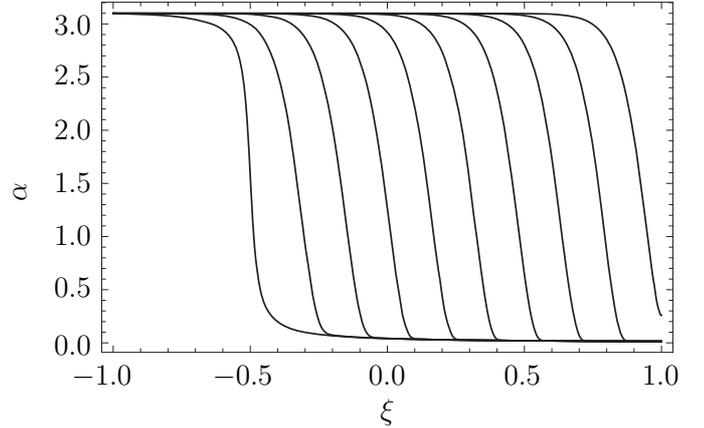}}
\caption{\label{fig7-3} 
FEM solution of Eq.\ 
(\ref{diffeq_2}) with $w=1.0$ and $v=2.0$.
The initial configuration is 
$\alpha(\xi,\tau=0)=(\pi/2)-\arctan(50x-25)$.  The subsequent 
evolution of the profile is shown at intervals of $\Delta\tau=0.05$.
The velocity of the profiles,
corresponding to the solitary wave Eq.\ (\ref{soliton_3}), 
fulfills Eq.\ (\ref{velocity}) in the main text. 
}
\end{figure}

 We show in Fig.\ \ref{fig7-2}  a  
 finite-element (FEM) solution to Eq.\ (\ref{diffeq_2}) 
for an initial $\arctan$-profile with two steps.  The figure shows the
initial profile of $\alpha$ versus $\xi$ and subsequent profiles at regular
time intervals.  
We observe that the initial profile progressively 
deforms into a stable profile that moves at constant speed.  This is 
the solitary wave $\alpha_s$ as given by Eq.\ (\ref{soliton_3}). 
Indeed, the initial step to the left in Fig.\ \ref{fig7-2}  
moves with larger velocity than that to the right, 
eventually overtaking and absorbing it. This is clear from Eq.\ 
(\ref{velocity}), which shows that the velocity of a solitary wave is 
proportional to the average $\alpha$ to the left and to the right of it. 
This is generally true: As long as $\alpha$ to the far left is larger than
the $\alpha$ to the far right, all deformations will be absorbed into a 
single solitary wave moving with the velocity given by Eq.\ 
(\ref{velocity}). In other words, the solitary wave 
exhibits a topological stability.  In Fig.\ \ref{fig7-3}  we show how an initial single-step  configuration rapidly evolves into the stable solitary wave $\alpha_s$.

	In conclusion we have shown that the combined effects of non-linear 
convection and diffusion for the dynamics of a wetting film  can lead to 
a rather intriguing  physics which, in our case, can be analyzed exactly. 
The local dynamical equation used describes a balance between 
viscous and capillary forces at the interface between the 
wetting film and a bulk liquid. When the bulk flow rate drops 
to zero along a pore, perturbations of the film spread diffusively 
and this non-linear diffusivity may,  as we have seen, lead to 
the appearance of shock fronts and, remarkably, to anomalous diffusion.
 In the presence of non-linear convection in addition to non-linear 
diffusion, we have seen the appearance of solitary waves. 
Numerical analysis has shown that the presence of several 
wave fronts eventually merges into a single solitary wave, the 
dynamical origin of which we have outlined. 
Concerning potential applications we, e.g., notice that
drainage processes, where a less wetting fluid displaces a 
more wetting fluid in a porous medium, appears to 
be fairly well understood today. This is, however, not the case 
in the opposite situation which we describe, i.e. 
imbibition, where a more wetting fluid displaces a less 
wetting fluid. We have also observed that the dynamical 
equation considered also enters in the context of non-linear heat 
conduction  where solitary waves therefore also may 
appear if a suitable form of convection is present.

This work has been supported by the Norwegian Research Council, EMGS AS,
Numerical Rocks AS and StatoilHydro AS. One of the authors (B.-S.S.) 
wishes to thank Professor Frederik G.\ Scholtz for a generous hospitality 
during a joint NITheP and Stias, Stellenbosch (S.A.), workshop in 2009 when the 
present work was in progress. The authors are grateful to referees for pointing out Refs.\cite{tsallis_96,liu_90}. During the submission process of the present paper a related work has appeared \cite{andrade_2010} with, however, a different focus then  the issues discussed here.


\end{document}